\documentclass[twocolumn]{aastex61}

\usepackage{graphicx}	
\usepackage{amsmath}	
\usepackage{amssymb}	
\usepackage{enumitem}

\newcommand{\Msun}{\,{\rm M_\odot}}
\newcommand{\Mblack}{M_\bullet}

\newcommand\mybar{\kern1pt\rule[-\dp\strutbox]{.8pt}{\baselineskip}\kern1pt}

\setlist[itemize]{noitemsep, topsep=0pt, leftmargin=*}

\shorttitle{Separating accretion and mergers in the cosmic growth of black holes}
\shortauthors{Pacucci \& Loeb}

\begin{document}

\title{Separating accretion and mergers in the cosmic growth of black holes \\ with X-ray and gravitational wave observations}

\correspondingauthor{Fabio Pacucci}
\email{fabio.pacucci@cfa.harvard.edu}

\author[0000-0001-9879-7780]{Fabio Pacucci}
\affil{Black Hole Initiative, Harvard University,
Cambridge, MA 02138, USA}
\affil{Center for Astrophysics $\vert$ Harvard \& Smithsonian,
Cambridge, MA 02138, USA}

\author[0000-0003-4330-287X]{Abraham Loeb}
\affil{Black Hole Initiative, Harvard University,
Cambridge, MA 02138, USA}
\affil{Center for Astrophysics $\vert$ Harvard \& Smithsonian,
Cambridge, MA 02138, USA}

\begin{abstract}
Black holes across a broad range of masses play a key role in the evolution of galaxies. The initial seeds of black holes formed at $z \sim 30$ and grew over cosmic time by gas accretion and mergers.
Using observational data for quasars and theoretical models for the hierarchical assembly of dark matter halos, we study the relative importance of gas accretion and mergers for black hole growth, as a function of redshift ($0<z<10$) and black hole mass ($ 10^3 \Msun <\Mblack < 10^{10} \Msun$).
We find that (i) growth by accretion is dominant in a large fraction of the parameter space, especially at $\Mblack > 10^8 \Msun$ and $z>6$; and (ii) growth by mergers is dominant at $\Mblack < 10^5 \Msun$ and $z>5.5$, and at $\Mblack > 10^8 \Msun$ and $z<2$. As the growth channel has direct implications for the black hole spin (with gas accretion leading to higher spin values), we test our model against $\sim 20$ robust spin measurements available thus far. As expected, the spin tends to decline toward the merger-dominated regime, thereby supporting our model.
The next generation of X-ray and gravitational-wave observatories (e.g. Lynx, AXIS, Athena and LISA) will map out populations of black holes up to very high redshift ($z\sim 20)$, covering the parameter space investigated here in almost its entirety. Their data will be instrumental to providing a clear picture of how black holes grew across cosmic time.

\end{abstract}


\section{Introduction} \label{sec:intro}
General relativity describes the gravitational field of astrophysical black holes in terms of two parameters only: mass and spin. Despite this simplicity, the effects of black holes on galaxy evolution are complex and far-reaching. They are ubiquitous and span a very wide range of masses, from stellar-mass ($\sim 10 \Msun$) to supermassive ($\sim 10^{10} \Msun$). Most galaxies host a massive black hole at their center and significant correlations exist between the mass of the compact object and some properties of the host, such as stellar mass or velocity dispersion \citep{Ferrarese_Merritt_2000, Gebhardt_2000, Kormendy_Ho_2013}. These relations indicate a profound relationship between the central black holes and their host galaxies.

In order to reach such a variety of masses spanning $\sim 9$ orders of magnitude, the initial population of black holes, formed at $z \sim 20-30$ (e.g., \citealt{BL01}), needs to grow over cosmic time, by gas accretion and mergers.
Growth by gas accretion can occur via a stable disk or chaotic accretion processes (e.g., \citealt{Shakura_Sunyaev_1976, King_2006}). The amount of mass added to the black hole depends on the mass-to-energy conversion efficiency factor $\epsilon$, as a fraction $\epsilon \dot{M} c^2$ ($c$ is the speed of light) of the mass influx per unit time ($\dot{M}=dM/dt$) is radiated away. Mergers between black holes also contribute to their growth over cosmic time (e.g., \citealt{Volonteri_2005_spin}), with a small fraction of the total mass ($\lesssim 10\%$, \citealt{Healy_2014}) being radiated away via gravitational waves (GWs).

The growth of black holes also directly affects the value of their spin. As the infalling gas generally possesses angular momentum, its accretion modifies the spin amplitude of the black holes (and also the orientation via the Lense-Thirring effect for already spinning black holes, see, e.g. \citealt{Bardeen_1975}).
In addition to gas accretion, black hole mergers can significantly modify the spin of the interacting black holes. Early heuristic arguments, valid for low mass ratios $q= (M_{\bullet 2}/M_{\bullet 1}) \ll 1$, where $M_{\bullet 1}$ and $M_{\bullet 2}$ are the masses of the two merging black holes, show that mergers typically spin black holes down \citep{Hughes_Blandford_2003, Volonteri_2005_spin}. Developments in numerical relativity (e.g., \citealt{Pretorius_2005}) have extended these results to major mergers ($q\sim 1$).
Despite significant advancements, the cosmological coevolution of black hole mass and spin is still an open problem \citep{Gammie_2004, Shapiro_2005, Berti_Volonteri_2008, Barausse_2012, Volonteri_2013_spin}.

Furthermore, the evolution of black hole mass and spin are strictly related to each other. On one hand, the spin directly influences how efficiently a black hole accretes mass, as it determines the mass-to-energy conversion efficiency factor $\epsilon$ \citep{Bardeen_1970, Novikov_Thorne}.
On the other hand, the spin of a black hole is determined by the mass it gained via accretion and mergers \citep{Berti_Volonteri_2008} and regulates how much energy can be extracted from the ergosphere of a black hole \citep{Blandford_Znajek_1977}.

In what follows, we use observational data and theoretical models to study the contribution of accretion and mergers for black hole growth over cosmic time. We then use $\sim 20$ spin measurements available from the literature at $z<1$ to test our model.
Future electromagnetic (e.g., Lynx, AXIS and Athena) and GW (LISA) observatories will be able to explore most of the parameter space in redshift and mass described here. These observatories will then provide invaluable constraints for black hole seeding models and cosmic growth. Our calculations use the latest values of the cosmological parameters \citep{Planck_2018}.

\section{Growth Model} 
\label{sec:theory}
We start with a summary of the black hole growth model employed in this study. As a preliminary step, we describe how we initialize galaxies at $z=30$ with black hole seeds. Then, we detail our assumptions to calculate the mass added by accretion and mergers.

\subsection{Black Hole Seeding Models} 
\label{subsec:seeds}
Assumptions on black hole seeding are needed in order to model their cosmological evolution. The formation of seeds directly from the first population of stars (Population III) is the most straightforward scenario. This simple model is challenged by the discovery of quasars, or active supermassive black holes with $\Mblack > 10^9 \Msun$, already in place at $z>6$ (e.g., \citealt{Fan_2006, Wu_2015, Banados_2018}).
Currently, the farthest quasar is detected at $z \approx 7.54$ \citep{Banados_2018}, only $\sim 700 \, \mathrm{Myr}$ after the Big Bang. Assuming that seeding occurred at $z \sim 30$, the formation of this quasar would require a continuous accretion process at the Eddington rate on a seed with minimum mass $\sim 10^3 \Msun$.
Questions concerning the rapid growth of black holes were already pointed out by \cite{Turner_1991} with the discovery of the first quasars at $4<z<5$ and became more pressing with the detection of $z > 6$ sources \citep{Haiman_Loeb_2001, Valiante_2017, Inayoshi_review_2019, Pacucci_Loeb_2019, Pacucci_Loeb_2019_mirage}. These constraints led to the development of formation models for heavy seeds, in an effort to decrease the growth time. Alternatively, several studies have pointed out that the growth time can also be decreased by super-Eddington phases of accretion (e.g., \citealt{Madau_2014_super, PVF_2015, Volonteri_2015, Lupi_2016, Pezzulli_2016, Begelman_Volonteri_2017, Regan_2019}).

Following \cite{Pacucci_2018}, we employ a bimodal distribution of $z=30$ black hole seeds. Heavy seeds are modeled as direct collapse black holes (DCBHs; e.g., \citealt{Bromm_Loeb_2003, Lodato_Natarajan_2006}) and light seeds as Population III stellar remnants (e.g., \citealt{Hirano_2014}).
The initial mass function for DCBHs is well described by a log-gaussian distribution with mean $\mu = 5.1$ and standard deviation $\sigma = 0.2$ \citep{Ferrara_2014}.
Population III stars of mass $m_{\star}$ are modeled with a low-mass cutoff $m_{c} = 10\, \mathrm{\Msun}$ and a Salpeter-like \citep{Salpeter_1955} exponent:
\begin{equation}
\Phi(\mathrm{Pop III},m_{\star}) \propto m_{\star}^{-2.35} \exp \left({-\frac{m_{c}}{m_{\star}}} \right) \, .
\end{equation}
To obtain the mass of black holes formed from this population of stars, we convolve this progenitor mass function with the relation \citep{Woosley_2002} between the stellar mass and the remnant mass.

While our fiducial model assumes that Population III remnants and DCBHs are representative formation channels for light and heavy seeds, respectively, we acknowledge that additional formation channels have been proposed. For example, black hole seeds heavier than Population III remnants can be formed by stellar collisions (e.g., \citealt{Devecchi_2009, Devecchi_2010, Devecchi_2012, Katz_2015, Boekholt_2018}) and black hole mergers (e.g., \citealt{Davies_2011, Lupi_2014}).

\subsection{Mass Growth Rate by Accretion} 
\label{subsec:m_a}
A black hole of mass $\Mblack$ fueled by a mass influx $\dot{M} = dM/dt$ gains mass at a rate
\begin{equation}
    \frac{d\Mblack}{dt} = (1-\epsilon)\dot{M} \, ,
\end{equation}
where the matter-to-energy conversion efficiency factor $\epsilon$ is customarily $\sim 10\%$ for radiatively efficient accretion disks \citep{Shakura_Sunyaev_1976}.
The remaining fraction of mass is radiated away, following the equation $L = \epsilon \dot{M}c^2$, where $L$ is the bolometric luminosity of the accreting source.
Introducing the Eddington ratio $f_{\mathrm{Edd}}$ to parameterize the accretion rate in terms of the Eddington rate ($\dot{M}_{\mathrm{Edd}}$)
\begin{equation}
    f_{\mathrm{Edd}} = \frac{\dot{M}}{\dot{M}_{\mathrm{Edd}}} \, ,
\end{equation}
the bolometric luminosity radiated by an accreting black hole can be expressed as
\begin{equation}
    L = \epsilon f_{\mathrm{Edd}} \dot{M}_{\mathrm{Edd}}c^2 \, ,
    \label{eq:luminosity}
\end{equation}
with the Eddington rate depending on the black hole mass and on the matter-to-energy conversion efficiency factor $\epsilon$
\begin{equation}
    \dot{M}_{\mathrm{Edd}} \approx \frac{2.2\times 10^{-9}}{\epsilon} \left(\frac{\Mblack}{\Msun}\right) \, [\mathrm{\Msun \, yr^{-1}}] = {\cal C} \frac{\Mblack}{\epsilon} \, .
\label{eq:M_edd}
\end{equation}
In Eq. (\ref{eq:M_edd}) we labeled the numerical factor as ${\cal C}$.
From the observed number density of quasars at different cosmic epochs, and some assumptions on $f_{\mathrm{Edd}}$ and $\epsilon$, it is thus possible to estimate the mass added by black holes from luminous accretion, employing Eq. (\ref{eq:luminosity}).

Early works (e.g., \citealt{Soltan_1982}) already made the argument that the total mass accreted by black holes over cosmic time can be calculated from their energy output, assuming that accretion is the ultimate source of the energy produced by quasars.
Following \cite{Soltan_1982}, the energy $E(L,t)$ produced in 1 Gpc$^3$ of comoving volume by sources with bolometric luminosity $L$ at the cosmic epoch $t$ can be inferred from the quasar luminosity function (LF) $\phi(L,z)$ as:
\begin{equation}
    E(L,t)dLdt = L \phi(L,z)dLdt \, .
\end{equation}

We employ a double power-law shape for the quasar LF, which well describes the number counts of quasars in a very wide redshift range (e.g., \citealt{Hopkins_2007, Jiang_2016, Yang_2016, Manti_2017}):
\begin{equation}
\phi ( L , z ) = \frac { \phi _ { \star } / L _ { \star } ( z ) } { \left[ L / L _ { \star } ( z ) \right] ^ { \alpha }  + \left[ L / L _ { \star } ( z ) \right] ^ { \beta } } \, .
\end{equation}
Here, $\phi _ { \star }$ is the normalization, $L _ { \star }$ is the break luminosity, $\alpha$ and $\beta$ are the slopes of the bright and of the faint ends, respectively. 
The values of the LF parameters for $0 \lesssim z \lesssim 7$ are taken from \cite{Jiang_2016}, \cite{Matsuoka_2018}, and \cite{Shen_2020}, while the values for $7 \lesssim z \lesssim 10$ are extrapolated from values at lower redshift, comparing our results with \cite{Manti_2017} whenever possible.

With assumptions on the function $L= {\cal F} (\Mblack)$, which ultimately requires a distribution of Eddington ratios (see below), it is possible to map the LF to the mass function $\varphi(\Mblack)$, see e.g. \cite{Brandon_2012}.

The mass added by accretion per unit time and per unit of comoving volume, $\dot{\rho}_{\rm a}$, for a black hole with original mass $\mu$ in the bin $[\Mblack, \Mblack + \Delta\Mblack]$ is then:
\begin{equation}
    \dot{\rho}_{\rm a} = \frac{(1-\epsilon)}{\epsilon} f_{\rm Edd} {\cal C} \int_{\Mblack}^{\Mblack+\Delta\Mblack} \varphi(\Mblack)\Mblack d\mathrm{\mu} \, .
\label{eq:M_acc}
\end{equation}
For all our calculations we assume $\epsilon=0.1$.
Equation \ref{eq:M_acc} is obtained assuming a generic accretion rate $f_{\rm Edd} \dot{M}_{\rm Edd}$ and that, at a given redshift $z$, the Eddington ratio is a function of the black hole mass only.

The distribution of Eddington ratios $f_{\rm Edd}$ as a function of $\Mblack$ and $z$ is highly uncertain and loosely constrained observationally, especially at high-$z$ (see e.g. \citealt{Fan_2019}). We estimate the function $f_{\rm Edd}(\Mblack,z)$ for $z\gtrsim 4.75$ from \cite{DeGraf_2012_Edd}, which uses the cosmological simulation \textsc{Massive Black}. For lower redshifts we use the results from \cite{Weigel_2017} and from the Sloan Digital Sky Survey data \citep{Shen_2011}, and interpolate for intermediate redshift.

There is an important caveat in this methodology for estimating $\dot{\rho}_{\rm a}$: it does not account for the mass added by quasars that are not counted in the LF. In other words, the presence of additional populations of quasars that are undetected because of obscuration \citep{Comastri_2015} or inefficient accretion \citep{Pacucci_MaxMass_2017, Pacucci_2017} cannot be taken into account here. Hence, our calculation of the mass added by accretion provides a lower limit.

\subsection{Mass Growth Rate by Mergers}

To estimate the mass added by mergers per unit time and comoving volume, $\dot{\rho}_{\rm m}$, we employ a semianalytical approach using a merger tree.
The code was originally developed for studies on warm dark matter \citep{Dayal_2017} and it is based on the well known approach described in \cite{Parkinson_2008}. The code is employed to track the evolution of a population of dark matter halos. We assume that the distribution follows a Sheth-Tormen halo mass function \citep{ST_1999}. Along with prescriptions for (i) black hole seeding, (ii) close pair formation probability for the black holes to merge, and (iii) energy released via GWs, this code can be used to calculate merger rates and, thus, mass added by mergers.

We sample logarithmically the cosmological scale factor $a = (z+1)^{-1}$ between $z=30$ and $z=0$, and assign one seed to each $z=30$ halo with a host mass $M_h \gtrsim 10^7 \Msun$. This limit on the mass of the host is equivalent to requiring that its virial temperature is above the atomic cooling threshold (e.g. \citealt{BL01}). The formation of heavy seeds ($\Mblack > 10^4 \Msun$) is a much rarer event than the formation of light seeds ($\Mblack < 10^3 \Msun$), as the physical requirements for the former channel are much more stringent (see, e.g., \citealt{Bromm_Loeb_2003, Latif_2013, Latif_2013b} and Sec. \ref{subsec:seeds} for additional details on seeding models). Analytical estimates show that heavy seeds form preferentially in more massive halos, with a distribution peaked around $M_h \sim 5 \times 10^7 \Msun$ \citep{Ferrara_2014}.
In our model, the ratio of heavy to light seeds is $\sim 1:100$, similar to the relative abundance of sources in the high-luminosity and the low-luminosity ends of the quasar LF (e.g. \citealt{Masters_2012, Shen_2020}). This ratio of light to heavy seeds reproduces the $z \sim 0$ quasar LF for $L_{\rm bol} \gtrsim 10^{43} \, \mathrm{erg \, s^{-1}}$ \citep{Hopkins_2007, Ricarte_2018}.
Once formed, black hole seeds accrete mass with the same distribution of Eddington ratios $f_{\rm Edd}$ described in Sec. \ref{subsec:m_a}.

\begin{figure}[h]
\includegraphics[angle=0,width=0.49\textwidth]{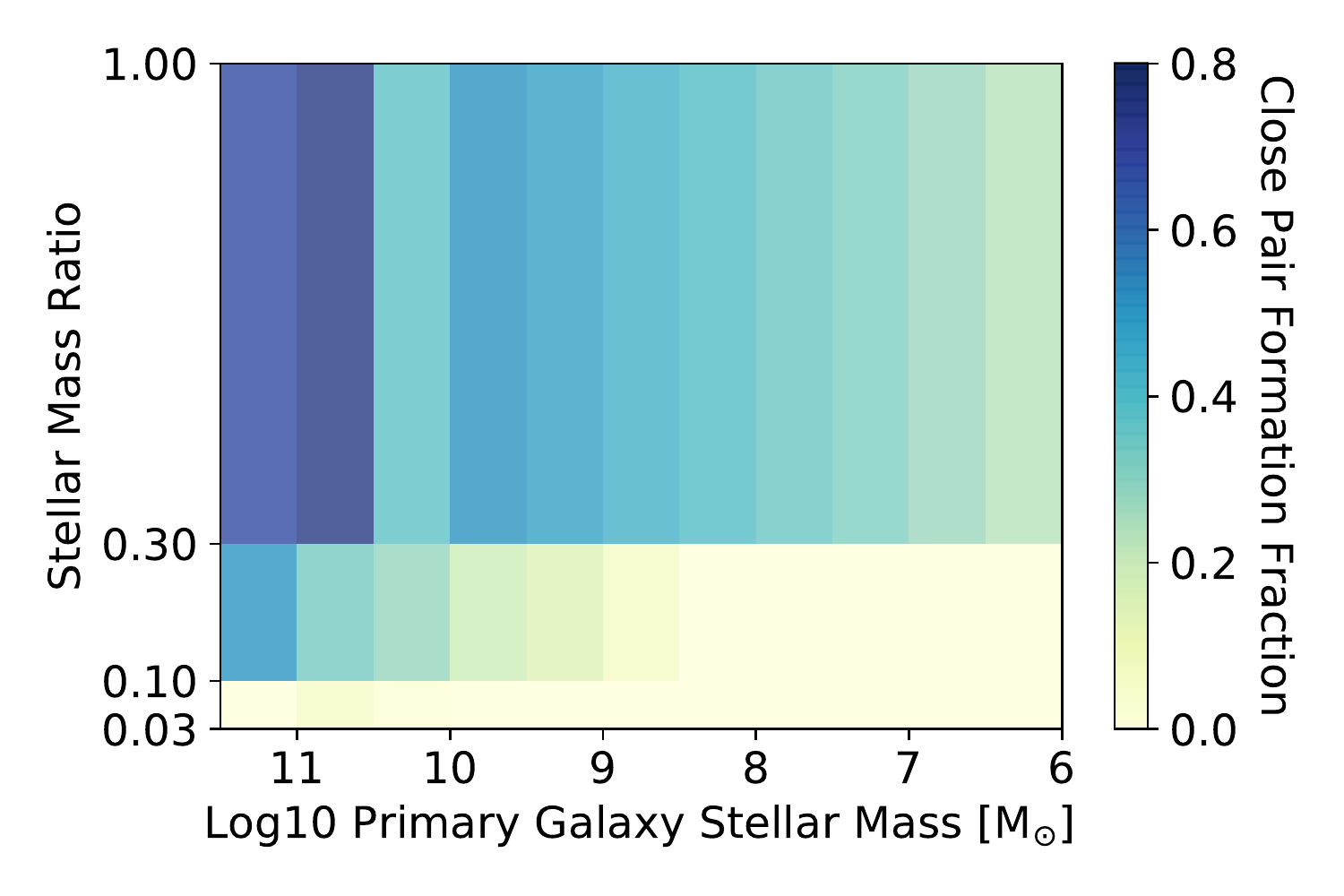}
\caption{Close pair formation fractions as a function of the stellar mass of the host and of the stellar mass ratio of the interacting galaxies. This plot is obtained by linearly extrapolating to lower stellar masses the results in \cite{Tremmel_2018}. Contrarily to the original study, we chose to represent the vertical axis in linear scale, to show that the highest stellar mass ratio bin ($m_{\star 2}/m_{\star 1}>0.3$) contains most of the range considered.}
\label{fig:heatmap}
\end{figure}

As two halos merge, their black holes form a close pair after a time delay equal to the dynamical friction time. Estimating the time for close pair formation is subject of active research and its accurate implementation is beyond the scope of this paper. In this study, we employ estimates of the probability of close pair formation derived in \cite{Tremmel_2018} from the cosmological simulation \textsc{Romulus}25. This probability does not directly depend on the black hole mass, but only on the stellar masses of the hosts. Due to the hierarchical seeding described above and in Sec. \ref{subsec:seeds}, heavier black holes are, on average, found in more massive hosts, hence stellar mass loosely tracks black hole mass. The model by \cite{Tremmel_2018} is valid down to $\Mblack \sim 10^6 \Msun$. Hence, for lighter black holes we linearly extrapolate their results, using the primary host stellar mass as extrapolation variable. Figure \ref{fig:heatmap} shows the values of the close pair formation fraction adopted in this paper, as a function of the stellar mass of the host and of the stellar mass ratio of the interacting galaxies.
The decreasing trend of the close pair formation probability with the mass is justified by the fact that the dynamical friction time increases when the mass of the black holes involved decreases \citep{Chandra_1943}. Although approximate, our approach offers an improvement when compared to previous semianalytical works (e.g., \citealt{Ricarte_2018}), which employ a mass-independent, fixed probability of close pair formation. Our approach, especially in the merger-dominated region at $z>5.5$ (see Sec. \ref{sec:results}), is also justified by \cite{Kulkarni_2012} which shows that at these cosmic epochs mergers are very frequent, with typical halo merger timescales significantly shorter than the Hubble time, for halo masses spanning four orders of magnitude.
It is also worth noting that the decrease, by a factor $\sim$few, of the probability of close pair formation for lighter black holes is counterbalanced by the fact that they are significantly more numerous than heavier black holes (by a factor $\sim 100$).

Considering two merging black holes with mass $M_{\bullet 1}$ and $M_{\bullet 2}$, we assume without loss of generality that $M_{\bullet 1} > M_{\bullet 2}$. Then, we define the mass added in a single merger as 
\begin{equation}
    M_{\rm single} = (1-\xi)\times M_{\bullet 2} \, .
\label{eq:add_single}
\end{equation}
Here, $0<\xi<1$ is the fraction of mass loss in gravitational radiation. The value of $\xi$ for a particular merger is computed via numerical relativity and depends on several parameters, such as the masses $M_{\bullet 1}$ and $M_{\bullet 2}$, their spins as well as their orientation. As we do not track most of these parameters, we fix $\xi = 0.11$ for all mergers. This value is an upper limit and valid for equal mass, aligned and maximally spinning binaries \citep{Healy_2014}. As in the case of the mass added by accretion (see Sec. \ref{subsec:m_a}), our calculation of $\dot{\rho}_{\rm m}$ is thus also a lower limit. We note, however, that fixing $\xi = 0.11$ affects our calculation of the mass added by mergers within $\sim 10\%$. As such, our results are not significantly sensitive to this choice. 

In order to calculate the merger rates for all black holes in our trees, we save for each merger the (i) redshift of the event and (ii) masses $M_{\bullet 1}$ and $M_{\bullet 2}$. Hence, we calculate the merger rate per unit comoving volume and unit redshift, $d^2N/d\mathrm{z}d\mathrm{V_c}$.
Then, we compute the merger rate per unit comoving volume and unit observed time by changing variables, as:
\begin{equation}
    \frac{d^2N}{d\mathrm{V_c} d\mathrm{t_{obs}}}= \frac{d^2N}{d\mathrm{z}d\mathrm{V_c}} \frac{dz}{d\mathrm{t_{rest}}(1+z)} \, ,
\end{equation}
where we have substituted $d\mathrm{t_{obs}} = d\mathrm{t_{rest}}(1+z)$ and we use standard cosmology for the derivative $dz/d\mathrm{t_{rest}}$.

With knowledge on the merger rates per unit comoving volume and unit observed time and with the prescription provided by Eq. (\ref{eq:add_single}), it is straightforward to compute the mass added by mergers per unit of time and of comoving volume, $\dot{\rho}_{\rm m}$.

\section{RESULTS} 
\label{sec:results}

We begin by discussing the mass added by accretion and mergers separately, for $1<z<6$. Then, we describe the relative importance of these growth modes at different redshifts and for different seeding models. Finally, we focus on the local universe ($z<1$) and discuss implications for the spin evolution.

\begin{figure*}[h]
    \includegraphics[width=1\textwidth]{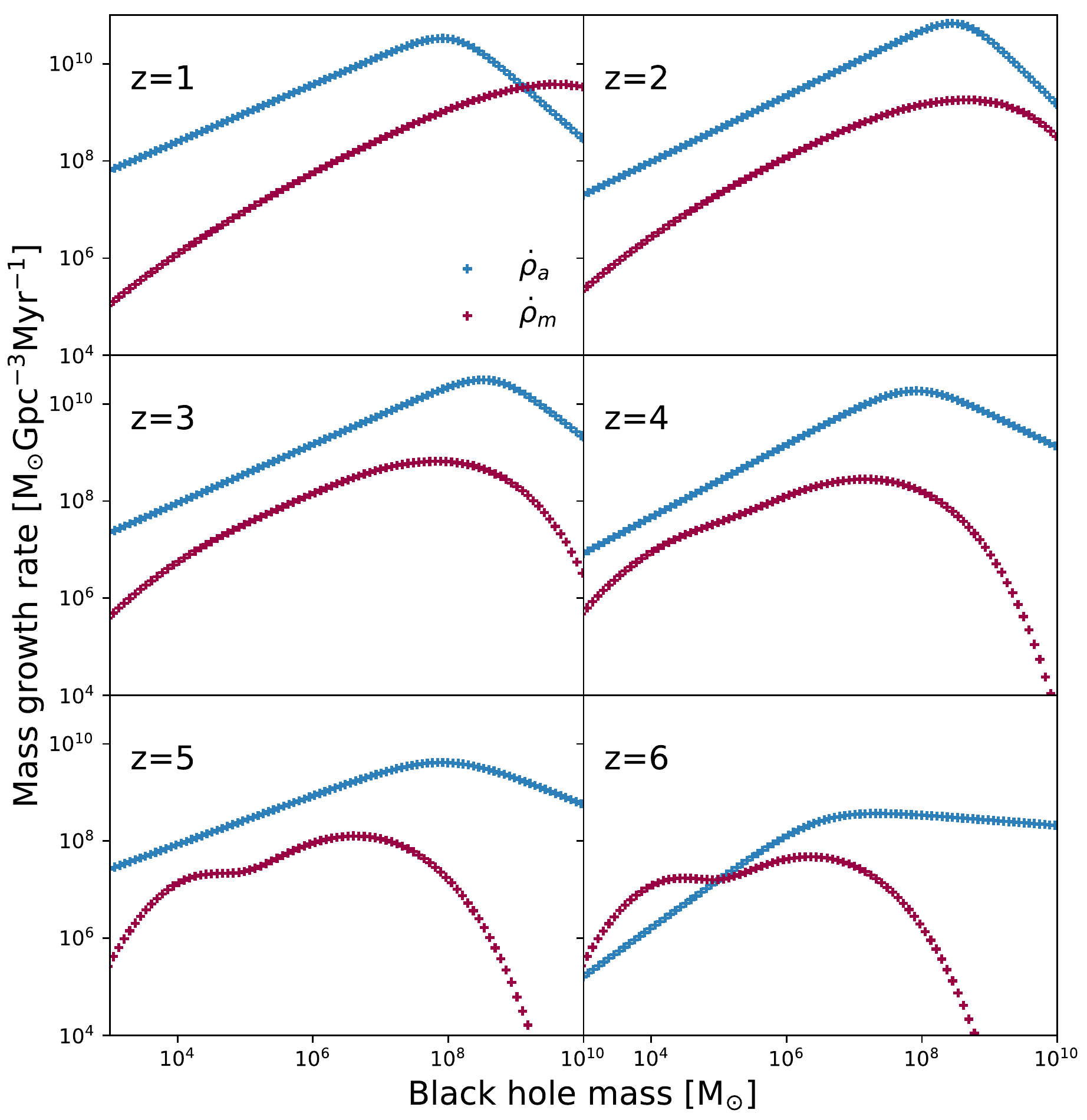} 
\caption{Mass added by accretion and mergers as a function of black hole mass, assuming that seeding at $z=30$ occurs with both light and heavy seeds. The six panels show our results for $z=1,2,3,4,5,6$, as indicated.}
\label{fig:multi}
\end{figure*}

\subsection{Accretion vs. Mergers for $1<z<6$}
In Fig. \ref{fig:multi} we show the mass growth rate by accretion (blue symbols) and by mergers (red symbols) as a function of black hole mass. We show these results in six panels for $z=1,2,3,4,5,6$, i.e. redshifts at which we have reliable measurements for the quasar LF.

These results, which include a mix of heavy and light black hole seeds at $z=30$, suggest that gas accretion is the primary growth mode for black holes. This is valid for almost every black hole mass and cosmic epoch shown. Exceptions are (i) black holes with $\Mblack\gtrsim 10^9 \Msun$ at $z=1$, and (ii) black holes with $\Mblack\lesssim 10^5 \Msun$ at $z=6$. The first exception can be explained by the fact that at $z\sim 1$ the amount of cold gas available for accretion starts to decline \citep{Power_2010} and may not be sufficient to fuel the most massive black holes ($\Mblack \gtrsim 10^9 \Msun$) at the Eddington rate. Conversely, at $z \sim 6$ the density of light seeds can be so high to drive up the mergers of small black holes and overcome the amount of mass that can be added by gas accretion. This explanation is also supported by the fact that, as $\dot{M}_{\rm Edd} \propto \Mblack$, for small black holes the absolute value of mass that can be added by gas accretion can be quite low. In particular, a black hole would need $\sim 50 \, \mathrm{Myr}$ of constant accretion at the Eddington rate to only double its mass, assuming a matter-to-energy conversion factor $\epsilon = 0.1$.

\subsection{Ratio of Mass Growth Rates by Accretion \\ and Mergers Over Cosmic History}

\begin{figure*}[h]
\includegraphics[angle=0,width=1.0\textwidth]{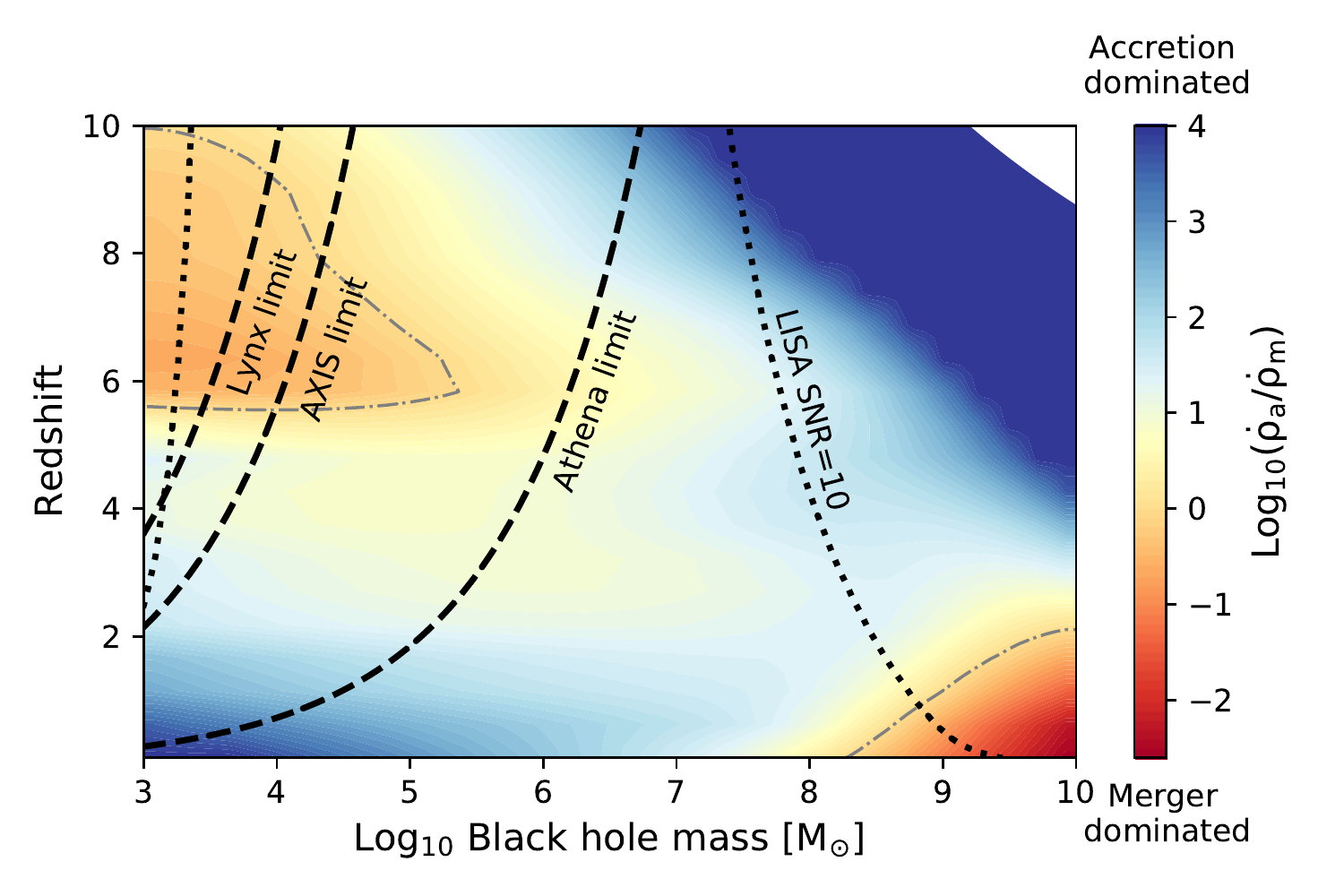}
\caption{Ratio ${\cal R} = \mathrm{Log}_{10}(\dot{\rho}_a/\dot{\rho}_m)$ between mass growth rates by gas accretion and mergers. The value of ${\cal R}$ is shown in the colorbar, as a function of black hole mass ($10^3 \Msun<\Mblack <10^{10} \Msun$) and redshift ($0<z<10$). The contour ${\cal R}=0$ is represented with a dashed-dotted line. The limits for Athena and for the prospective Lynx and AXIS X-ray observatories are shown with a thick, dashed line. In both cases, detectable sources extend below the lines. The limit corresponding to a signal-to-noise ratio of $10$ for the GW observatory LISA is shown with the two thick, dotted lines. This plot assumes that both heavy and light seeds are formed at $z=30$. The whited-out area on the top right corner indicates the region of the parameter space $\Mblack-z$ where we predict no black holes (see the main text for details).} 
\label{fig:history}
\end{figure*}

In order to compare the importance of gas accretion and mergers over cosmic history, we define the logarithm of the ratio of the mass growth rate in the two modes,
\begin{equation}
    {\cal R} = \mathrm{Log}_{10}\frac{\dot{\rho}_a}{\dot{\rho}_m} \, .
\end{equation}

Figure \ref{fig:history} shows the evolution of the parameter ${\cal R}$ for masses $10^3 \Msun<\Mblack<10^{10} \Msun$ and in the range $0<z<10$.
In this contour plot, the locus where ${\cal R} = 0$ (i.e., where gas accretion and mergers contribute equally) is shown as a dashed-dotted line.
This plot confirms that growth by accretion is dominant over a large range of masses and redshifts. Moreover, it also shows more clearly the regions (see Fig. \ref{fig:multi}) where, instead, growth by mergers is dominant: (i) at low masses ($\Mblack < 10^5 \Msun$) and high redshift ($z > 5.5$), and (ii) at high masses ($\Mblack > 10^8 \Msun$) and low redshift ($z < 2$).

\begin{figure*}
\includegraphics[angle=0,width=0.5\textwidth]{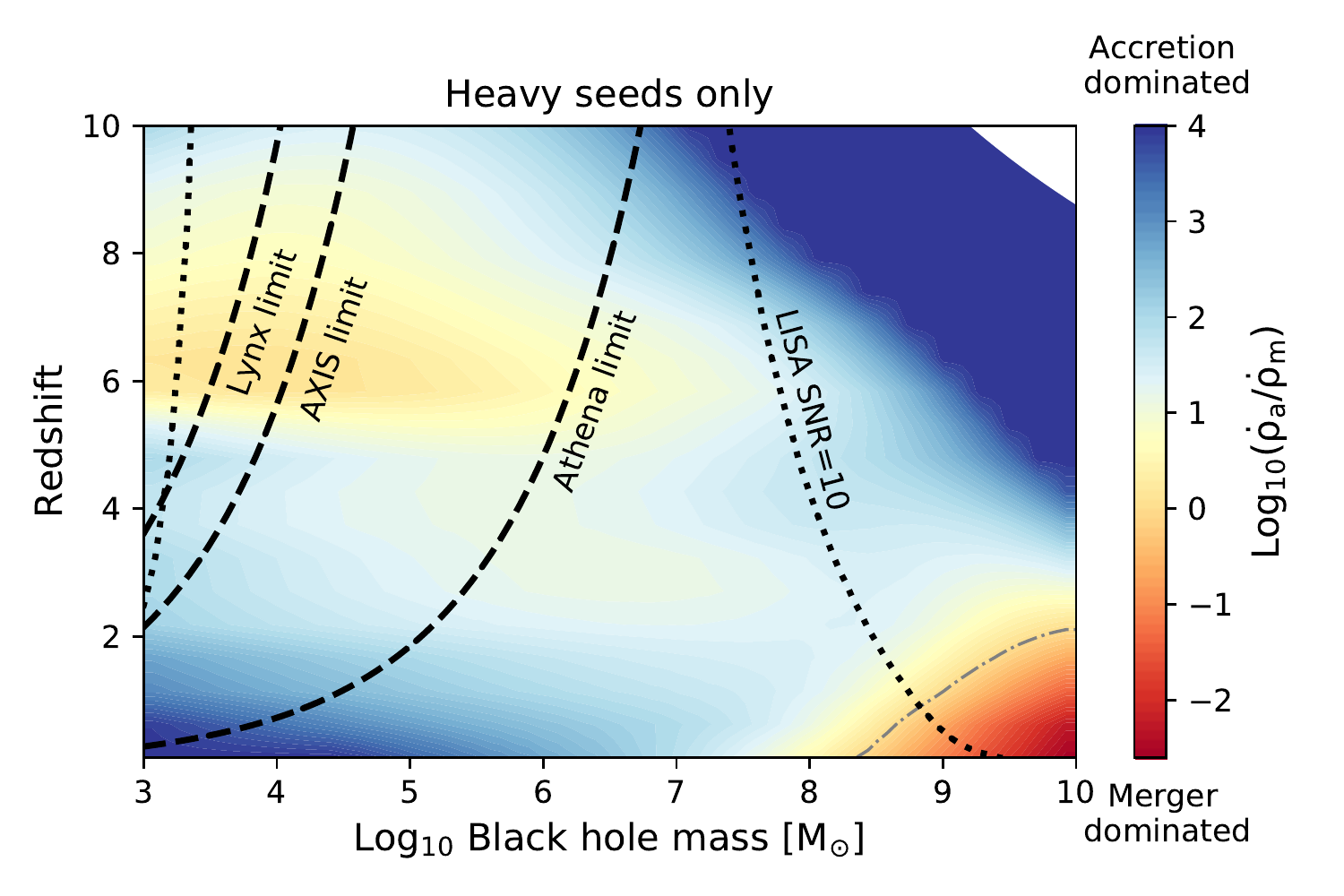}
\includegraphics[angle=0,width=0.5\textwidth]{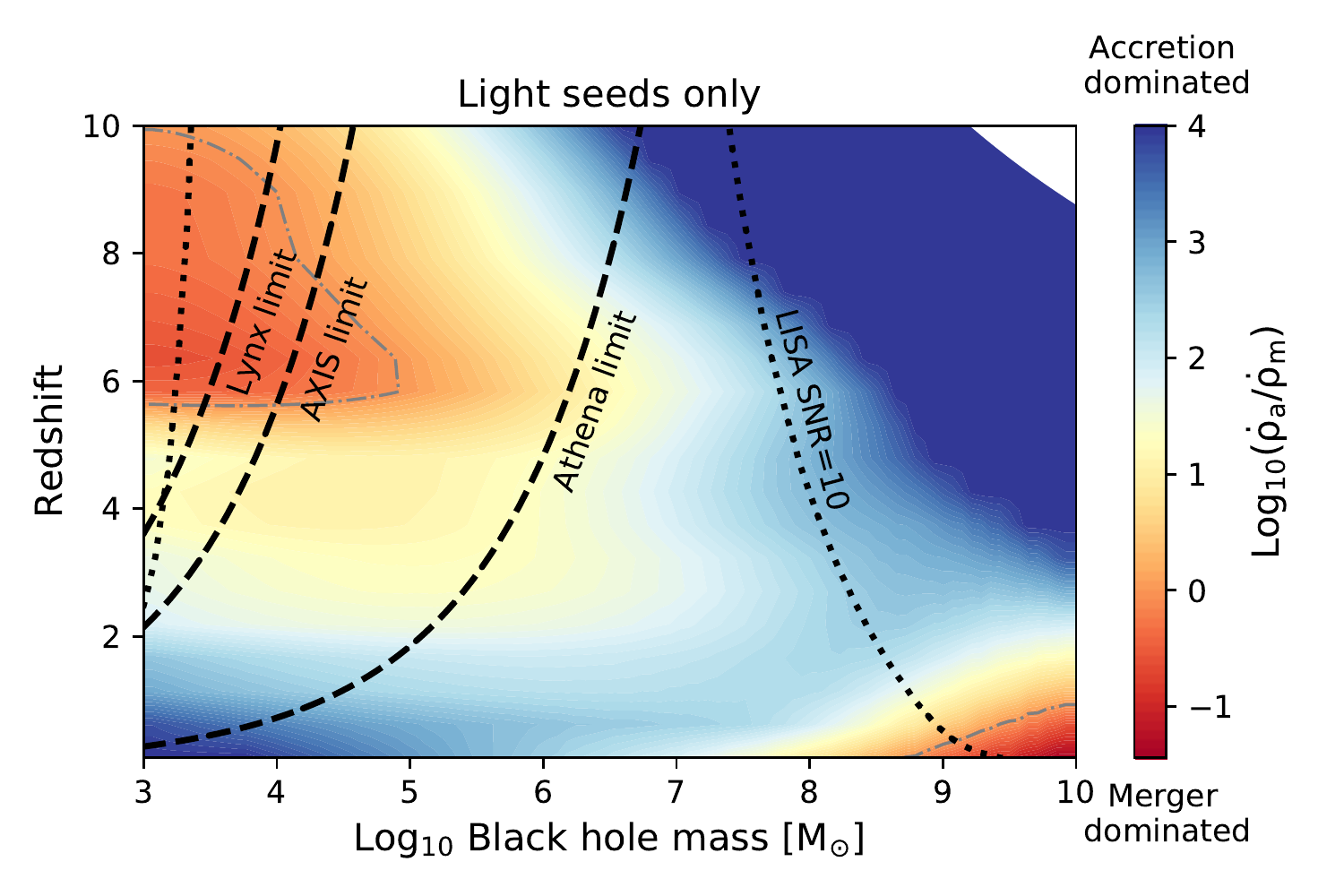}
\caption{Same as in Fig. \ref{fig:history}, but with different assumptions for seed formation at $z=30$. \textbf{Left panel:} only heavy seeds with initial masses in the range $10^4 \Msun<\Mblack<10^6 \Msun$ and log-gaussian distribution with mean $\mathrm{Log}_{10}\Mblack[\Msun] = 5.1$ are formed. \textbf{Right panel:} only light seeds with initial masses $\Mblack <10^3 \Msun$ and Salpeter-like distribution of Population III progenitors are formed. See Sec. \ref{subsec:seeds} for a detailed description of seeding models.}
\label{fig:history_seeding}
\end{figure*}

Figure \ref{fig:history} shows an additional, remarkable information: at high redshift ($z\gtrsim 5$) and high masses ($\Mblack \gtrsim 10^8 \Msun$) growth by accretion becomes significantly dominant, by $\gtrsim 4$ orders of magnitude. This is due to the fact that, at such early epochs, very massive black holes have not formed yet, or they are extremely rare. Hence, the probability of growth by mergers for these objects is very low. Note that, as the mass growth by mergers becomes very small, ${\cal R}$ becomes very large in this region. To improve the visualization of the color scale, we introduce an upper bound ${\cal R} = 4$ in Figs. \ref{fig:history} and \ref{fig:history_seeding}.

On the top right side of the contour plot, the whited-out area indicates the region of the parameter space $\Mblack -z$ where we predict no black holes. This very loose upper limit is computed by assuming that, at a given $\Tilde{z}$, the most massive black hole theoretically achievable derives from the formation, at $z=30$, of a seed of $10^6 \Msun$ (i.e., a very massive, heavy seed) which continuously grows at the Eddington rate up to $\Tilde{z}$.

As shown by the detection limits for \href{https://wwwastro.msfc.nasa.gov/lynx/docs/LynxConceptStudy.pdf}{Lynx}, AXIS \citep{AXIS}, \href{https://www.cosmos.esa.int/documents/400752/400864/Athena+Mission+Proposal/18b4a058-5d43-4065-b135-7fe651307c46}{Athena} and LISA (\citealt{LISA_2017}, with a signal-to-noise ratio of 10), a combination of X-ray and GW observations can track almost the totality of this plane in redshift and black hole mass. The only undetectable sources will be at very low masses ($\Mblack \lesssim 10^{3.5} \Msun$) and high redshift ($z\gtrsim 4$). This additional section of the parameter space could be covered in the future by third-generation, ground-based GW detectors, such as the Einstein Telescope \citep{Sesana_2009}.

The results for ${\cal R}(m,z)$ shown in Fig. \ref{fig:history} can be well fitted for $z<6$ (i.e., the region where a reliable quasar LF is available) with a cubic surface. Defining $m = \mathrm{Log}_{10} \Mblack$ $[\Msun]$, we construct the fitting function as
\begin{equation}
\begin{split}
    {\cal R}(m,z) \approx & \, p_{00} + p_{10}m + p_{20}m^2 
    + p_{30}m^3 + \\
    & + p_{01}z + p_{02}z^2 + p_{03}z^3 + \\
    & + p_{12}mz^2 +p_{21}m^2z + p_{11}mz \, ,
\end{split}
\end{equation}
and we list its parameters in Table \ref{tab:parameter}.

\begin{table}[h]
\begin{center}
\begin{tabular}{ccccc}
\hline
\textbf{$p_{00}$} & \textbf{$p_{10}$} & \textbf{$p_{20}$} & \textbf{$p_{30}$} & \textbf{$p_{01}$} \\ \hline
4.8878           & -0.37770         & 0.063241          & -0.010222        & -1.1195          \\ \hline
\textbf{$p_{02}$} & \textbf{$p_{03}$} & \textbf{$p_{12}$} & \textbf{$p_{21}$} & \textbf{$p_{11}$} \\ \hline
0.35505          & -0.024092        & -0.019512        & 0.050761          & -0.28528         \\ \hline
\end{tabular}
\end{center}
\caption{List of parameter values for the cubic surface to fit ${\cal R}(m,z)$.}
\label{tab:parameter}
\end{table}

In Fig. \ref{fig:history_seeding} we show the same representation of Fig. \ref{fig:history}, but assuming that only heavy seeds (left panel) or light seeds (right panel) are formed at $z=30$.
In the case of heavy seeds only, we recognize that the importance of mergers for mass growth is significantly reduced: regions with ${\cal R} < 0$ are shrunk in size, especially at low masses and high redshift. This is explained by the fact that heavy seeds are much rarer ($\sim 1:100$ in our model) and the probability of merger events is thus significantly reduced.
On the other side of the seeding spectrum, with light seeds only, the importance of mergers at low masses and high redshift is reestablished. In addition, the high-redshift area where gas accretion is significantly dominant (${\cal R} \sim 10^4$) is slightly more extended. This is a consequence of the challenge in building very massive black holes ($\Mblack \gtrsim 10^8 \Msun$) at $z \gtrsim 6$ via mergers and with light seeds only. Thus, growth of heavy black holes can only occur via gas accretion at those redshifts.

\subsection{Gas Accretion and Mergers at $z<1$: \\ Data from Spin Measurements}

\begin{figure*}
\includegraphics[angle=0,width=1.0\textwidth]{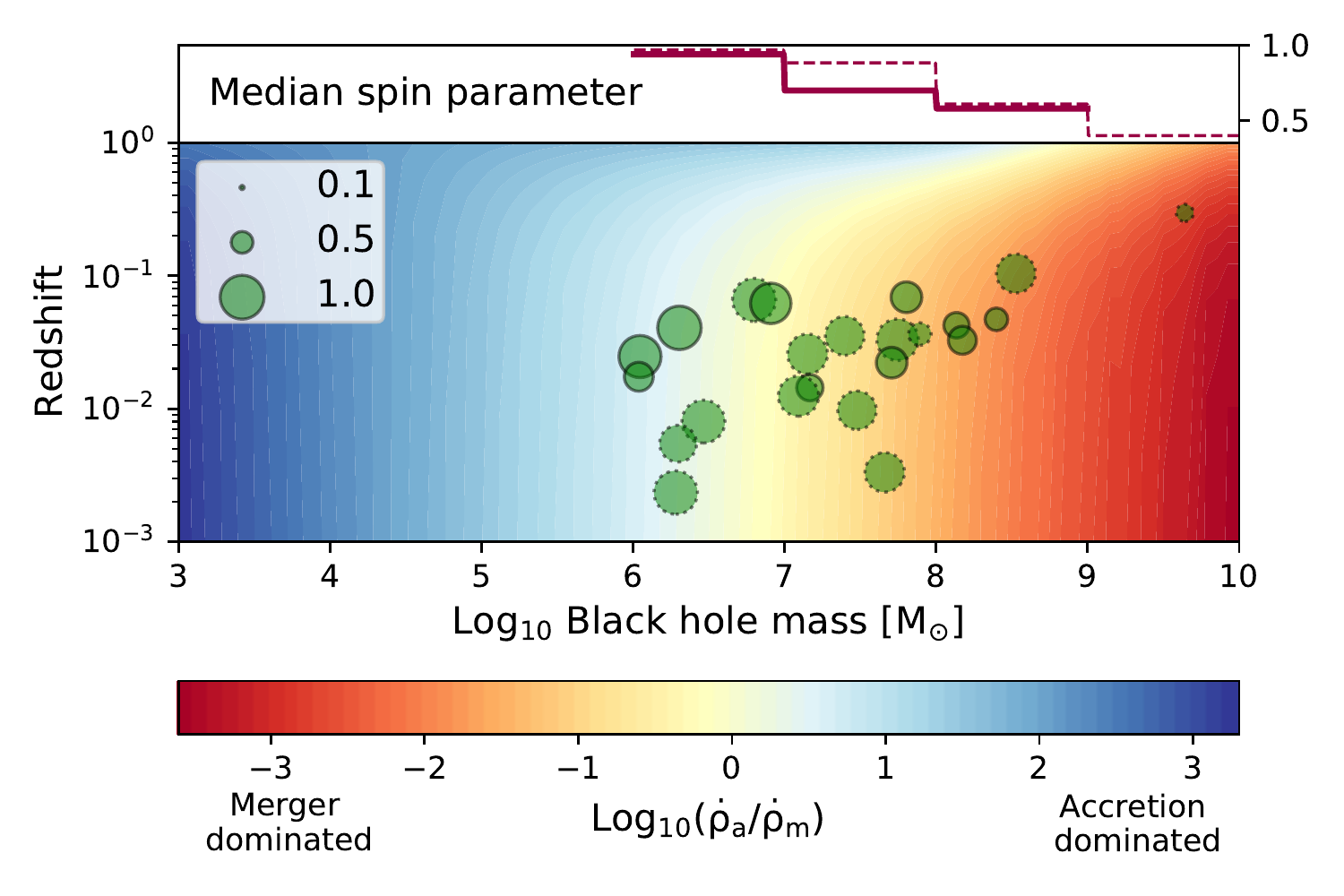}
\caption{Same as in Fig. \ref{fig:history}, but for $z<1$. The green circles indicate $23$ black holes with a robust measurement of spin \citep{Vasudevan_2016, Reynolds_2019}. The size of the circle indicates the magnitude of the spin parameter $a$ measured. The size scale for $a=0.1, 0.5, 1.0$ is shown in the legend. Circles with a dashed border point out that the spin measurement is a lower limit. In the top panel, the median spin parameter is calculated binning the data in black hole mass, with bin size equal to one dex. The solid line shows the median computed excluding lower limits, while the dashed line includes all points. The data show a significant decline of the median spin parameter for increasing black hole masses, thus supporting our model for the evolution of the parameter ${\cal R}$.}
\label{fig:spins}
\end{figure*}

To conclude, Fig. \ref{fig:spins} shows the evolution of the parameter ${\cal R}$ for $z<1$, assuming both light and heavy seeds at $z=30$. The contour plot shows the increasing dominance of growth by mergers at larger black hole masses (see also Fig. \ref{fig:history}). The spin of a black hole is significantly affected by its growth mode. Generally speaking, growth by accretion is predicted to lead to higher spins when compared to growth by mergers only (see, e.g., \citealt{Berti_Volonteri_2008, Barausse_2012} and our discussion in Sec. \ref{sec:disc_concl}). Measurements of black hole spins at $z<1$ \citep{Reynolds_2019} allow us to test our model. In the following, the spin is indicated with the dimensionless spin parameter $0<a<1$, where $a=0$ is a Schwarzschild (i.e., nonrotating) black hole, and $a=1$ is a maximally rotating black hole (although this value is not achievable in practice, \citealt{Thorne_1974}).

Figure \ref{fig:spins} shows the values of the spin parameter $a$ for $23$ black holes, obtained from \cite{Vasudevan_2016} and \cite{Reynolds_2019}. The top panel shows the median value of the spin calculated by binning the data in black hole mass, with $\Delta \mathrm{Log}_{10}(\Mblack) = 1$. Despite a paucity of accurate spin measurements available thus far, the data show a significant decline of the median spin parameter for increasing black hole masses, at $z<1$. As previous theoretical models \citep{Hughes_Blandford_2003, Gammie_2004, Pretorius_2005, Shapiro_2005, Berti_Volonteri_2008, Barausse_2012} indicate that the highest spin parameters ($a\sim 0.9$) are reached by gas accretion, the data support our model for the evolution of the parameter ${\cal R}$.

It is worth noting that these results are not significantly affected by our assumption of a fixed value of the mass-to-energy conversion efficiency factor $\epsilon=0.1$.
As mentioned in Sec. \ref{sec:intro}, the black hole spin directly affects the value of $\epsilon$: nonrotating black holes are characterized by an efficiency of $\sim 5.7\%$, while maximally rotating black holes by $\sim 32\%$. Rapidly spinning black holes are thus less efficient in growing via mass accretion, because $\dot{\rho}_a \propto (1-\epsilon)/\epsilon$, see Eq. \ref{eq:M_acc}. Hence, black holes accreting in the left side of Fig. \ref{fig:spins} should be characterized by larger values of $\epsilon$, as we expect them to be rapidly spinning. The assumption of constant matter-to-energy conversion efficiency factor $\epsilon=0.1$ throughout the mass range considered does not significantly affect our results for $\dot{\rho}_a$, and hence for the parameter ${\cal R}$. In fact, $\dot{\rho}_a \propto (1-\epsilon)/\epsilon$ (see Eq. \ref{eq:M_acc}) and the correction associated with a variable $\epsilon$ would be up to $\sim 1$ order of magnitude, while the parameter ${\cal R}$ varies by about $\sim 6$ orders of magnitude over the whole mass range. As we do not yet have a clear picture of how gas accretion and mergers affect the spin of black holes in the whole mass range considered, we prefer to keep the model simple and assume $\epsilon=0.1$ for $10^3 \Msun<\Mblack<10^{10} \Msun$.

\section{Discussion and Conclusions} 
\label{sec:disc_concl}

By employing observational data for the quasar LF and theoretical models for the hierarchical assembly of dark matter halos, we have investigated the mass added to black holes by gas accretion and mergers, as a function of redshift ($0<z<10$) and black hole mass ($ 10^3 \Msun <\Mblack < 10^{10} \Msun$). Our work is motivated by the crucial role of the growth mechanism in determining the only observables of astrophysical black holes: mass and spin.

Assuming a bimodal seeding at $z=30$ featuring both light ($< 10^3 \Msun$) and heavy ($\sim 10^5 \Msun$) seeds, and defining  the logarithm of the ratio between mass growth rate by gas accretion and mergers as ${\cal R} = \mathrm{Log}_{10}(\dot{\rho}_a/\dot{\rho}_m)$ our main findings are the following: 
\begin{itemize}
    \item growth by accretion is dominant in a large fraction of the parameter space, especially at $\Mblack > 10^8 \Msun$ and $z>6$ where it is significantly dominant, with ${\cal R} \sim 10^4$;
    \item growth by mergers is dominant, instead, at $\Mblack < 10^5 \Msun$ and $z>5.5$, and at $\Mblack > 10^8 \Msun$ and $z<2$.
\end{itemize}
We note that some recent papers (e.g., \citealt{Cowie_2020}) used X-ray data from the Chandra observatory to suggest a very low black hole accretion density at $z=5-10$. This might be due to the fact that early black hole populations are buried behind very large, Compton-thick, column densities of gas and become undetectable even in the X-rays (see, e.g., \citealt{Comastri_2015} for a corresponding discussion).

The dominance of accretion vs. mergers for black hole growth can be inferred indirectly through spin measurements.
The spin value reached by a black hole is significantly dependent on the mass accretion mode that was dominant during its growth. Several models and numerical relativity simulations \citep{Hughes_Blandford_2003, Gammie_2004, Pretorius_2005, Shapiro_2005, Berti_Volonteri_2008, Barausse_2012} strongly suggest that the highest spin parameters $a > 0.9$ are reached by gas accretion, while mergers would lead to more moderate values of $a \sim 0.5$. We showed that this general relation supports our picture of the dichotomy accretion vs. mergers at $z<1$, where our study predicts that higher-mass black holes ($\Mblack \gtrsim 10^7 \Msun$) accrete mostly via mergers. From the limited sample ($\sim 20$) of black holes with accurate spin measurements, we infer that higher-mass black holes are indeed characterized by lower values of spins \citep{Vasudevan_2016, Reynolds_2019}.

If the same models to map accretion mode to spin value are valid also at $z>1$, our study predicts that massive black holes at high-$z$ are characterized by spin parameters close to the maximal value. On the other hand, we predict that lower-mass black holes at high-$z$ are characterized by moderate values of the spin parameter.

Due to a more significant availability of cold gas at high-$z$, a fraction of black holes might be accreting at super-Eddington rates. Simple estimates presented in \cite{Begelman_Volonteri_2017} suggest that a fraction $\sim 10^{-3}$ of active galactic nuclei could be accreting at super-Eddington rates already at $z\sim 1$, and possibly a fraction $\sim 10^{-2}$ at $z\sim 2$. Extrapolating this trend to higher redshift, we expect the majority of black holes to be accreting close to, or above the Eddington rate at $z\gtrsim 5$. Accretion at super-Eddington rates might not spin up the black holes in the same manner as accretion at lower-Eddington rates. Future numerical simulations will be fundamental in mapping out the relation between accretion mode and spin at higher redshift and in properly translating our Fig. \ref{fig:history} to predictions for black hole spin.

The vast majority of sources of interest for our study will be observable by next-generation X-ray and GW observatories. In particular, we showed that Athena, AXIS, Lynx and LISA will be instrumental in investigating the properties of such sources.
Many methods exist for measuring the mass of black holes, and they employ either estimates derived from the dynamics of gas/stars accelerated by the gravitational field of the black hole (e.g., masers, gas/stellar dynamics, reverberation mapping) or indirect relations with some properties of the hosts (see, e.g., \citealt{Peterson_2014} for an extensive review). Measuring spins is much more challenging, as it requires probing very close to the event horizon ($R_S = 2G\Mblack/c^2$, where $G$ is the gravitational constant), the region significantly affected by the rotation of the black hole. The iron line fluorescence method is the most extensively employed technique and is based on gravitational redshift of atomic features in the X-ray spectrum, at the energy $E_{\rm Fe} \sim 6.4 \, \mathrm{keV}$ \citep{Reynolds_1997}. Additional techniques rely on thermal continuum fitting to probe regions of the accretion disk close to the innermost stable circular orbit, as its extension is directly related to the spin value \citep{Bardeen_1970, Novikov_Thorne, Narayan_McClintock_2013}. Currently there are robust spin measurements for $\lesssim 30$ super-massive black holes \citep{Vasudevan_2016, Reynolds_2019}.
These early measurements suggest that most black holes below $3\times 10^7 \Msun$ are spinning very fast ($a\gtrsim 0.9$), while heavier black holes typically show lower spin values $a\sim 0.5-0.7$ \citep{Reynolds_2019}. This is in agreement with the expectation from our work at $z<1$.

Next-generation observatories will map the broad population of black holes out to high redshifts and be instrumental in providing a clear picture of how black holes grew across cosmic time.

\vspace{0.3cm}
We thank the anonymous referee for constructive comments on the manuscript. F.P. acknowledges fruitful discussions with Ramesh Narayan, Michael Tremmel and Richard Mushotzky.
This work was supported by the Black Hole Initiative at Harvard University, which is funded by grants from the John Templeton Foundation and the Gordon and Betty Moore Foundation.





\bibliographystyle{mnras}
\bibliography{ms}
\label{lastpage}
\end{document}